\documentclass[aps,preprintnumbers,amsmath,amssymb,prd,superscriptaddress,longbibliography,twocolumn]{revtex4-2}

\usepackage{dcolumn}% Align table columns on decimal point
\usepackage{color}
\usepackage{subfigure}  % use for side-by-side figures
\usepackage{feynmp-auto}
\usepackage{pgfplots}
\def\comment#1{}

\newcommand{\nc}{\newcommand}
\nc{\beq}{\begin{eqnarray}}
\nc{\eeq}{\end{eqnarray}}
\nc{\scs}{\scriptstyle}
\nc{\setval}{\fmfset{wiggly_len}{3mm} \fmfset{arrow_len}{1.5mm}
\fmfset{arrow_ang}{13} \fmfset{dash_len}{1.5mm}\fmfpen{0.125mm}
\fmfset{dot_size}{2thick}}

\usepackage{bm,latexsym,mathrsfs,enumerate,color}
\usepackage[mathcal]{euscript}
\usepackage[breaklinks=true,unicode=true,urlcolor = blue,colorlinks = true,citecolor = blue,linkcolor = blue]{hyperref}
\usepackage{graphicx}
\usepackage{wrapfig}
\usepackage{dsfont}
\usepackage{bbm}
\usepackage[percent]{overpic}
\usepackage{empheq}
\usepackage{tikz}

\usepackage{outlines}
\usepackage[capitalize]{cleveref}
\usetikzlibrary{positioning,shapes.misc,calc,3d,arrows,decorations.pathmorphing,decorations.markings,arrows.meta,external,tqft}
\usepackage{tikz-3dplot}
\renewcommand{\vec}[1]{\bm{#1}}

\def\slashchar#1{\setbox0=\hbox{$#1$}           % set a box for #1
\dimen0=\wd0                                 % and get its size
\setbox1=\hbox{/} \dimen1=\wd1               % get size of /
\ifdim\dimen0>\dimen1                        % #1 is bigger
\rlap{\hbox to \dimen0{\hfil/\hfil}}      % so center / in box
#1                                        % and print #1
\else                                        % / is bigger
\rlap{\hbox to \dimen1{\hfil$#1$\hfil}}   % so center #1
/                                         % and print /
\fi}                                         %

\DeclareMathAlphabet\mathbfcal{OMS}{cmsy}{b}{n}

\usepackage{mathtools}% http://ctan.org/pkg/mathtools
\usepackage{graphicx}
\usepackage{dcolumn}
\usepackage{amssymb}
\usepackage{amsmath}
\usepackage{slashed}
\usepackage{bm}
\usepackage{bbm}
\usetikzlibrary{external}
\begin{document}

\title{String-like Theory of Quantum Hall Interfaces}
\author{O\u{g}uz T\"urker}
\author{Kun Yang}
\affiliation{National High Magnetic Field Laboratory and Department of Physics,
Florida State University, Tallahassee, Florida 32306, USA}
%\pacs{73.43.Nq, 73.43.-f}

\begin{abstract}

We derive the effective theories for quantum hall droplets with attractive interaction among the constituent particles. In the absence of confining potentials such droplets are defined by their freely moving interfaces (or boundaries) with the vacuum. We demonstrate the effective theories take forms similar to string theories. Generalization to interfaces between different quantum Hall liquids is discussed.

\end{abstract}

\date{\today}

\maketitle

\section{Introduction}
Quantum Hall (QH) effect is one of the most active research fields in condensed matter physics\cite{BOOK}. There are many different approaches to describe QH physics \cite{Ezawa2013,Stone1992,Chakraborty1995,Sarma1996,Yoshioka2002,Prange2012,Jain2007}. In the \emph{bulk}, the low energy effective theory of a QH liquid is the Chern-Simons (CS) theory  \cite{Wen2007}. At \emph{edge}, the low energy effective theory consists of two pieces, the topological and the dynamical piece \cite{Wen2007}. The topological piece is dictated by the bulk CS theory (known as bulk-edge correspondence) so that the overall theory describing the bulk and the edge is gauge-invariant. Just like the bulk CS theory it describes the degrees of freedom at the edge (usually chiral bosonic and Majorana fermionic modes), but contains no dynamical information. Dynamics at the edge comes from external confining potential, which determines the location of the edge and dictates the edge Hamiltonian\cite{BOOK,Wen2007}.	

Recently, one of us considered a different type of boundaries \cite{Kun2021}, namely \emph{the quantum Hall interfaces} that separate different QH phases, whose locations are {\em not} determined (or pinned) by any external potential. A special class of such interfaces separate a QH liquid from vacuum, which (superficially) look very much like edges. Instead of repulsive interaction between the electrons in the system, Ref. \cite{Kun2021} invoked attractive interactions to hold the QH liquid together, thus eliminating the necessity of the external confining potential. Accordingly, such a quantum Hall droplet (QHD) is formed spontaneously and can move freely. In the absence of confining potential the interface behaves like a string, whose dynamics is dominated by the string tension\cite{Kun2021}.

Ref. \cite{Kun2021} focused on the low-energy excitation spectra of various interfaces, which can be accessed easily by modifying the Hamiltonian of the corresponding edge theories. To describe the general interface dynamics however, we need to study large distortions of the interface configurations from the ground state. This is the task of our present work. As we will see this requires a reformulation of the entire theory (including the topological term), which leads to a string-like theory for the interfaces. At low-energy this allows for improvement of the results of Ref. \cite{Kun2021} by including curvature effects in a systematic way, which accounts for the (relatively small but definitely noticeable) discrepancy between theory prediction and numerical data. More importantly we demonstrate QH interfaces realize certain non-relativistic versions of both bosonic and superstrings; in the latter case there are Majorana fermion modes that live along the strings. This facilitates synergy between condensed matter and high-energy physics.

%In this work, we will show that  the \emph{curvature} of such interface has also  contributions to the low energy effective theory which leads to a closer predictions with the numerical calculations.  Thus we will study effects of the geometry of quantum Hall interfaces to  its dynamics. In addition, we   also derive a Nambu-Goto type action for such interface and discuss the reparametrization properties.
%the  As a consequence, this system is much more  for such system have  and while leaving topological piece  the edge of the   instead of confining the quantum Hall system.
%In this, we will study the effects of the geometry of such interface. and show that the curvature is   to  is a Instead, they consider an interface with suc  with en external electric field authors assume  a  o dynamic piea  , because   physics of a finite FQH system can   a well studied phenomenon.  In a recent work in ref \cite{Kun2021}, authors show that it is in theory possible to  have quantum Hall droplet (QHD) living in a trivial liquid such that it is not confined by an external potential. In this work we will study  the interface of this (QHD) with the trivial surrounding  liquid. We will derive and action defined on a world-sheet of the interface of (QHD) which is  invariant under non-relativistic diffeomorphisms (Diff) and Galilean transformations. We also discuss the relations of this system with Newton-Cartan geometry.

The rest of the paper is organized as follows. In Sec. \ref{sec:Interface between integer Quantum Hall liquid and vacuum} we start by including the curvature effects in the energy functional for the interface between fermionic $\nu=1$ QH liquid and vacuum, and demonstrate this leads to significant improvement in the comparison between first-principle prediction of the low-energy excitation spectrum and numerical results. In Sec. \ref{sec:Pfaffian-Vacuum interface} we show similar improvement for the Pfaffian-vacuum interface. Sec. \ref{sec: Chern-Simons theory with time-dependent boundary} is devoted to the derivation of the topological term in the action when the interface configuration deviates significantly from that of the ground state. Some concluding remarks are offered in Sec. \ref{sec: concluding remarks}, where we discuss possible generalization of our approach to interfaces between different quantum Hall liquids.

\section{Interface between fermionic $\nu=1$ Quantum Hall liquid and vacuum}
\label{sec:Interface between integer Quantum Hall liquid and vacuum}

We start with the simplest possible QH interface separating the fermionic $\nu=1$ QH liquid and vacuum, with the former stabilized by an {\em attractive} Haldane $V_1$ pseudopotential\cite{BOOK} and the Pauli principle. Due to the simplicity of this state parameters of the string-like theory can be determined from microscopic calculations.

\begin{figure}
\includegraphics{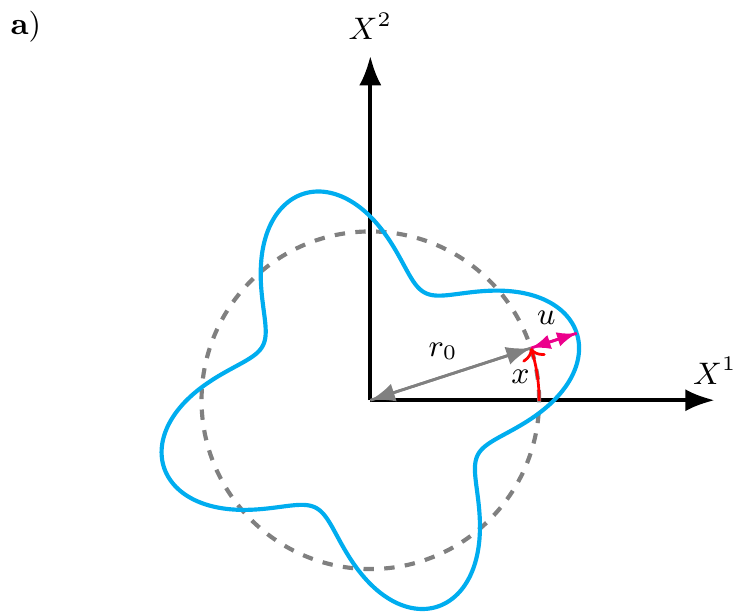}
\includegraphics{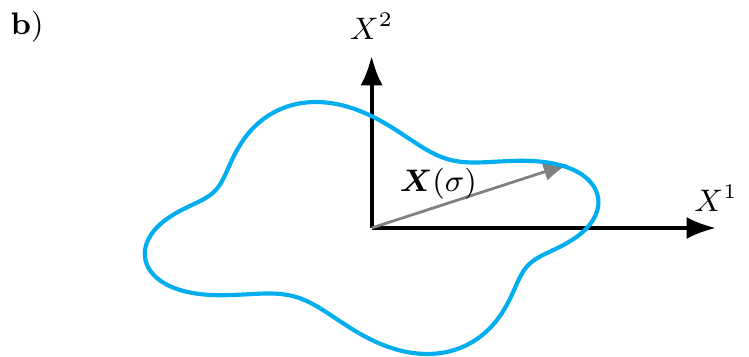}
	\caption{\label{fig:stringgauge} Illustration of a quantum Hall droplet with emphasis on its interface with vacuum. $\mathbf{a)}$ A configuration with infinitesimal distortion. Dashed circle with radius $r_0$ shows the equilibrium configuration, cyan is the distorted configuration, $x$ is the coordinate along the equilibrium, $u$ is the transversal displacement of the interface from the equilibrium and $X^{1,2}$ are the coordinate system. $\mathbf{b)}$ A configuration with large distortion. }
\end{figure}

The effective action of a droplet with $N$ fermions and radius $r_0=\sqrt{2N}$ (we set the magnetic length, $\ell$, to unity) in the low-energy (or linear) regime is a slight modification of the edge theory with a single chiral bosonic field\cite{Kun2021}:
\begin{equation}
	S=-\frac{1}{4\pi}\int dt \int_0^{2\pi r_0} dx\dot{\phi}\phi'-\frac{T_0}{2}\int dt \int_0^{2\pi r_0} dx(u')^2, \label{eq:originalmodel}
\end{equation}
where $T_0$, is the string tension, $x$ is the coordinate along the unperturbed string, $t$ is time, prime is  $\partial_x$, dot is  $\partial_t$ and $u$,  the transversal displacement from equilibrium (see \cref{fig:stringgauge}-a),  is given by
\begin{equation}
	\phi'=u,\label{eq:cons}
\end{equation}
where   $\phi$ is a compact scalar with identification
\begin{equation}
	\phi\sim\phi+2\pi.
\end{equation}
The first term in \cref{eq:originalmodel} is the aforementioned  topological piece  which emerges from the bulk CS and the second term, the  dynamical piece,  is the length \footnote{It is actually the difference of equilibrium length with the length of the distorted edge.} times the tension in the limit of  $|u'|\ll1$ and $r_0\to \infty$. We will use the term ``linear regime'' for such approximation, which is valid when the deviation to the ground state configuration is small and slowly varying.

\subsection{Action beyond the linear regime}

When we consider larger deformations, higher order terms in $u$ and its derivatives become important, and \cref{eq:originalmodel} needs to be modified. As a first step we have to replace the approximated arc length in \cref{eq:originalmodel} (that renders it quadratic) with the actual arc length.
%\cref{eq:originalmodel} needs to be modified. As a first step we have to replace the approximated arc length in \cref{eq:originalmodel} (that renders it quadratic) with the actual arc length. Thus we have
%	\begin{align}
%	S&=-\frac{1}{4\pi}\int d^2x\dot{\phi}\phi'-T_0\int d^2x\sqrt{\Big(\frac{u}{r_0}+1\Big)^2+(u')^2}{\notag}\\
%	 &+\int d^2x\lambda(\phi'- u){\notag}\\
%	 & +\int dtw\Bigg(\frac{1}{2}\int dxr_0\Big(\frac{u}{r_0}+1\Big)^2-\pi r_0^2\Bigg)\label{eq:originalmodel2},
%\end{align}
However, at this point the energy of the system is completely determined by the length of the interface. Thus classically energy conservation means conservation of  the string length. This together with the area conservation constraint (which comes from the incompressibility of the QHD) implies that the classical equation of motion of the string would be trivial, namely the only allowed motions are rigid spinning and motion of the center of mass. Thus, as we will see below such a modified action would be incomplete; it  misses the curvature dependence of the local energy density, which has significant effect on the dynamics.  %\textcolor{blue}{we may just remove this subsection removing it would not effect the flow.}
%\subsection{Non-constant Tension}

A convenient way to incorporate the effect of curvature is to make the tension $T$ a function of the curvature, $\kappa$, of the string which is given as \cite{Carmo1993}:
\begin{equation} \kappa=\frac{2\frac{u'^{2}}{r_0}-(1+\frac{u}{r_0})(u''-\frac{u}{r_0^2}-\frac{1}{r_0})}{\Big(\big(\frac{u}{r_0}+1\big)^2+u'^2\Big)^{\frac{3}{2}}}\label{eq:curvature}.
\end{equation}
In this case, Hamiltonian is no longer just proportional to the string length, thus energy conservation and area conservation no longer lead to trivial motion.
The reason we choose to make $T$ a function of curvature is for a planar curve (i.e. a curve lying on a plane) arc length and curvature completely determines a curve up to a rigid motion \cite{Carmo1993}\footnote{However, if we were in 3+1 dimensional space, \emph{torsion} (i.e. deviation of a curve from a planar curve) of the curve would be the another quantity along with curvature to completely characterize a curve \cite{Carmo1993}.}.
Thus our more general model that includes the curvature effect is
\begin{align}
	S&=-\frac{1}{4\pi}\int d^2x\dot{\phi}\phi'-\int d^2xT(\kappa)\sqrt{\Big(\frac{u}{r_0}+1\Big)^2+(u')^2}{\notag}\\
	&+\int d^2x\lambda(\phi'- u){\notag}\\
	& +\int dtw\Bigg(\frac{1}{2}\int dxr_0\Big(\frac{u}{r_0}+1\Big)^2-\pi r_0^2\Bigg)\label{eq:originalmodel3},
\end{align}
where $\lambda$ is the Lagrangian multiplier field to enforce \cref{eq:cons}, $w$ is also a Lagrangian multiplier field which only depends on time, and it enforces area conservation. If we set $r_0\to \infty$ and expand the \cref{eq:originalmodel3} up to quadratic order and set $T\to T_0$ we will get \cref{eq:originalmodel} back.

\subsection{Extracting the curvature-dependent tension}

We can find $T(\kappa)$ by extending the method of Ref. \cite{Kun2021} that determined $T(\kappa=0)$. In units of $|V_1|$ (set to be one from now on) the ground state energy of the droplet in terms of number of electrons was found to be \cite{Kun2021}
\begin{equation}
	\langle V \rangle=\frac{1}{2}N\left(4-\frac{4^{2-N}(2N-1)!}{(N-1)!N!}\right).
\end{equation}
Expanding at large $N$ yields
\begin{equation}
	\langle V\rangle=-2 N+\frac{4 \sqrt{N}}{\sqrt{\pi }}-\frac{\sqrt{\frac{1}{N}}}{2 \sqrt{\pi
	}}+\mathcal{O}\left(\left(\frac{1}{N}\right)^{3/2}\right). \label{eq:gsen}
\end{equation}
The first term is the bulk contribution, and the rest is the surface contribution that can be compared with our low energy effective theory \cref{eq:originalmodel3}. According to \cref{eq:originalmodel3} and \cref{eq:curvature}, the energy of a perfectly circular interface is just $2\pi r_0T(1/r_0)$.  As we show below this with \cref{eq:gsen} determines $T(\kappa)$ up to third order in $\kappa$:
\begin{equation}
	T(\kappa)=T_0+T_1\kappa+T_2\kappa^2+\mathcal{O}(\kappa^4),\label{eq:texp}
\end{equation}
where $T_0$, $T_1$ and $T_2$ are constants. Comparing  \cref{eq:texp} with \cref{eq:gsen} yields
\begin{equation}
	T\Big (  \kappa=\frac{1}{r_0}=\frac{1}{\sqrt{2N}}\Big)2\pi \sqrt{2N}=\frac{4 \sqrt{N}}{\sqrt{\pi }}-\frac{\sqrt{\frac{1}{N}}}{2 \sqrt{\pi
	}},
\end{equation}
which implies  $T_1=0$ and
\begin{equation}
	T_2=-\frac{T_0}{4}=-\frac{1}{\sqrt{8\pi^3}} \label{eq:coefs},
\end{equation}
%note that we have already set $\ell=1$.
Finally, we can substitute \cref{eq:coefs} to \cref{eq:texp} and find for an \emph{arbitrary local} curvature $\kappa$,
\begin{equation}
	T(\kappa)=\frac{4-\kappa^2}{\sqrt{8\pi^3}}+\mathcal{O}(\kappa^4).
\end{equation}
Note our finding $T_1 = 0$ is not an accident; in fact all odd terms in the expansion \cref{eq:texp} are zero, because if we exchange the regions of QH liquid and the vacuum the sign of the curvature flips, but the surface contribution to the energy remains same due to the particle-hole symmetry of the two-body Hamiltonian.

\subsection{Effects of the curvature dependent tension on the low-energy excitation spectrum}
\begin{figure}
	\scalebox{.3}{\includegraphics{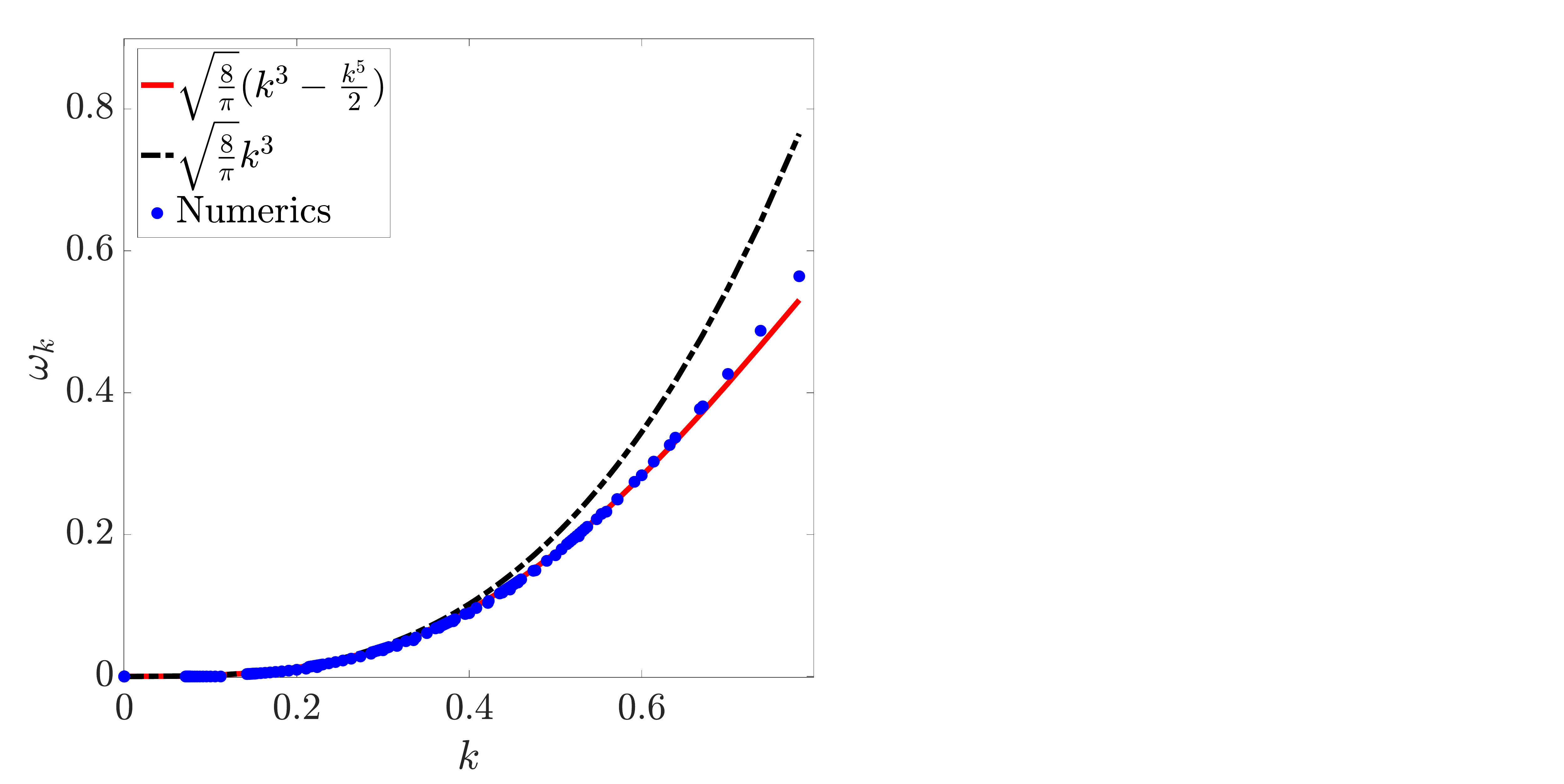}}
	\caption{Comparison between the bosonic spectrum (with and without curvature corrections) and numerical results of Ref. \cite{Kun2021}, for the interface between a $\nu=1$ fermionic integer quantum Hall liquid and the vacuum. The agreement is much better when the curvature effects are included (red solid line) than that without the curvature effects (black dashed line).}
\label{fig:newdisrel}
\end{figure}

We substitute \cref{eq:texp} to \cref{eq:originalmodel3} and expand it up to quadratic order in $u$ which gives
\begin{align}
	S&=-\frac{1}{4\pi}\int d^2x\dot{\phi}\phi'+\int d^2x\lambda(\phi'-u){\notag}\\
	&-\int d^2x\Bigg(T_0\frac{u'^2}{2}+T_2\bigg(\frac{u^2}{r_0^4}+\frac{4uu''+\frac{3}{2}u'^2}{r_0^2}+u''^2\bigg)\Bigg){\notag}\\
	&+\int dtw\Bigg(\frac{1}{2}\int dxr_0\Big(\frac{u}{r_0}+1\Big)^2-\pi r_0^2\Bigg)+\mathcal{O}(u^3)\label{eq:originalmodel4},
\end{align}
where we have dropped unimportant total derivative terms. %In order to have a better insight on the negligible terms  of \cref{eq:originalmodel4}
To obtain the mode spectrum we first integrate out the $\lambda$ and then express the action in term of $\phi$ in momentum-frequency space:
\begin{equation}
	\phi(t,x)\propto \int d\omega\sum_k e^{-i\omega t+ikx}\phi_{k\omega},
\end{equation}
where the discrete momentum (due to finite size of QHD) is $k=n/r_0$ with $n\in\mathbb{Z}$. The action in momentum-frequency space is thus
\begin{align}
	S&=\int_\omega\sum_k|\phi|^2k\bigg(\frac{\omega}{4\pi}-T_0\frac{k^3}{2}-T_2\Big(\frac{k}{r_0^4}-\frac{5}{2r_0^2}k^3+k^5\Big)\bigg){\notag}\\
	& +\int dtw\bigg(\frac{1}{2}\int dxr_0\Big(\frac{u}{r_0}+1\Big)^2-\pi r_0^2\bigg)\label{eq:originalmodel5}.
\end{align}
We can neglect terms of order of $1/r_0$ and higher for sufficiently large $N$. The quadratic action is thus (back to space-time domain)
\begin{align}
	S&=-\frac{1}{4\pi}\int d^2x\dot{\phi}\phi'-\int d^2x\Big[T_0\frac{u'^2}{2}+T_2u''^2\Big]{\notag}\\
	&+\int d^2x\lambda(\phi'-u) +\mathcal{O}(u^3)+\mathcal{O}\left(\frac{1}{r_0}\right)\label{eq:originalmodel6}.
\end{align}
Comparing to \cref{eq:originalmodel}, we find a correction due to $T_2$ in this limit. The chiral mode dispersion relation can be easily extracted:
\begin{equation}
	\omega=\sqrt{\frac{8}{\pi}}\Big(k^3-\frac{k^5}{2}\Big),\label{eq:cordisrel}
\end{equation}
which compares extremely well with numerical results (see \cref{fig:newdisrel}); the agreement is significantly better than the spectrum without taking into account the curvature effect.
 %If we compare \cref{eq:originalmodel6} with \cref{eq:originalmodel}, we see that at the linear regime the only contribution of a curvature dependent  tension is $T_2u''^2$.
We note if we had included in \cref{eq:texp} higher order terms, their contribution would vanish at the present level of approximation (i.e. keeping quadratic terms only and taking the $r_0\to\infty$ limit). This is clearer if we just express the curvature in the linear regime as
\begin{equation}
	\kappa=u''+\mathcal{O}(u^2)+\mathcal{O}\left(\frac{1}{r_0}\right).
\end{equation}
This means the spectrum \cref{eq:cordisrel} is {\em exact} in the large $N$ or $r_0$ limit; any further deviation with numerics (see \cref{fig:newdisrel}) will have to be accounted for by nonlinearity (or interaction among the bosons), and/or finite-size effect.

%If we explicitly include $\ell$ we have,
%\begin{equation}
%	\omega=\ell^3\sqrt{\frac{8}{\pi}}\Big(k^3-\ell^2\frac{k^5}{2}\Big),\label{eq:cordisrel2}
%\end{equation} and if we set $k=n/(\ell \sqrt{2N})$,
%\begin{equation}
%	\omega=\sqrt{\frac{8}{\pi}}\Bigg(\left(\frac{n}{ \sqrt{2N}}\right)^3-\frac{1}{2}\left(\frac{n}{\sqrt{2N}}\right)^5\Bigg),\label{eq:cordisrel2}
%\end{equation}

\subsection{General form of string action}
\label{subsec: General form of string acion}

Our discussion thus far has assumed the deviation of the interface from its ground state (circular) configuration is (still) small. As a result our action (\ref{eq:originalmodel3}) is an expansion of the deviation parameterized by $u$. However since there is no confinement potential, such deviation is unbounded in the present case of interface (and unlike the edge), rendering $u$ hard to define, let alone using it for expansion in general (see \cref{fig:stringgauge}-b). In the general case we should instead work directly with the world sheet of the of interface ${\vec{X}}(t, \sigma)$ where $\sigma$ parameterizes a point on the interface (see \cref{fig:stringgauge}-b). The general form of the interface action is a functional of the world sheet  ${\vec{X}}(t, \sigma)$, and is a sum of a topological (or Wess-Zumino-like) term and an energy term:
\begin{equation}
	S=S_{top}[{\vec{X}}(t, \sigma)] - \int{dt} E[{\vec{X}}(t, \sigma)],
\label{eq:string action}
\end{equation}
where, based on our analysis above,
\begin{equation}
E[{\vec{X}}(t, \sigma)] = \int{d\sigma} T(K(\sigma))|\partial{\vec{X}}/\partial\sigma|.
\end{equation}
The derivation of $S_{top}[{\vec{X}}(t, \sigma)]$ is subtler, and will be the subject of Sec. \ref{sec: Chern-Simons theory with time-dependent boundary}. What should be clear at this point is the action (\ref{eq:string action}) takes the form of some kind of (bosonic) string theory\cite{Polchinski1998}.

\section{Pfaffian-Vacuum interface}
\label{sec:Pfaffian-Vacuum interface}

In this section we consider a more complicated and interesting interface, that between a Moore-Read (Pfaffian) state formed by bosons of filling factor $\nu=1$ with a trivial vacuum. Here again the $\nu=1$ boson droplet is formed spontaneously by an {\em attractive} $V_0$ pseudopotential, supplemented by an infinitely strong three-body interaction that prevents the bosons from collapsing, which also makes the Moore-Read (Pfaffian) wave function the exact ground state \cite{Kun2021}. The edge theory of the regular Moore-Read state consists of a chiral Majorana mode and a chiral bosonic mode which are decoupled and both have linear dispersions. In the present interface case however, the dispersion of the bosonic mode was found to have a similar form as in the previous section, while the dispersion of the Majorana mode remains linear\cite{Kun2021}. We can still expect them to decouple, since by a simple dimensional analysis (or power counting) there is no relevant interaction term between the bosonic and fermionic modes. While we can not extract parameters from microscopics in this case as in the previous section, we expect the curvature effects to be present and affect the spectrum of the bosonic mode, which we now analyze.

\begin{figure}
	\scalebox{.3}{\includegraphics{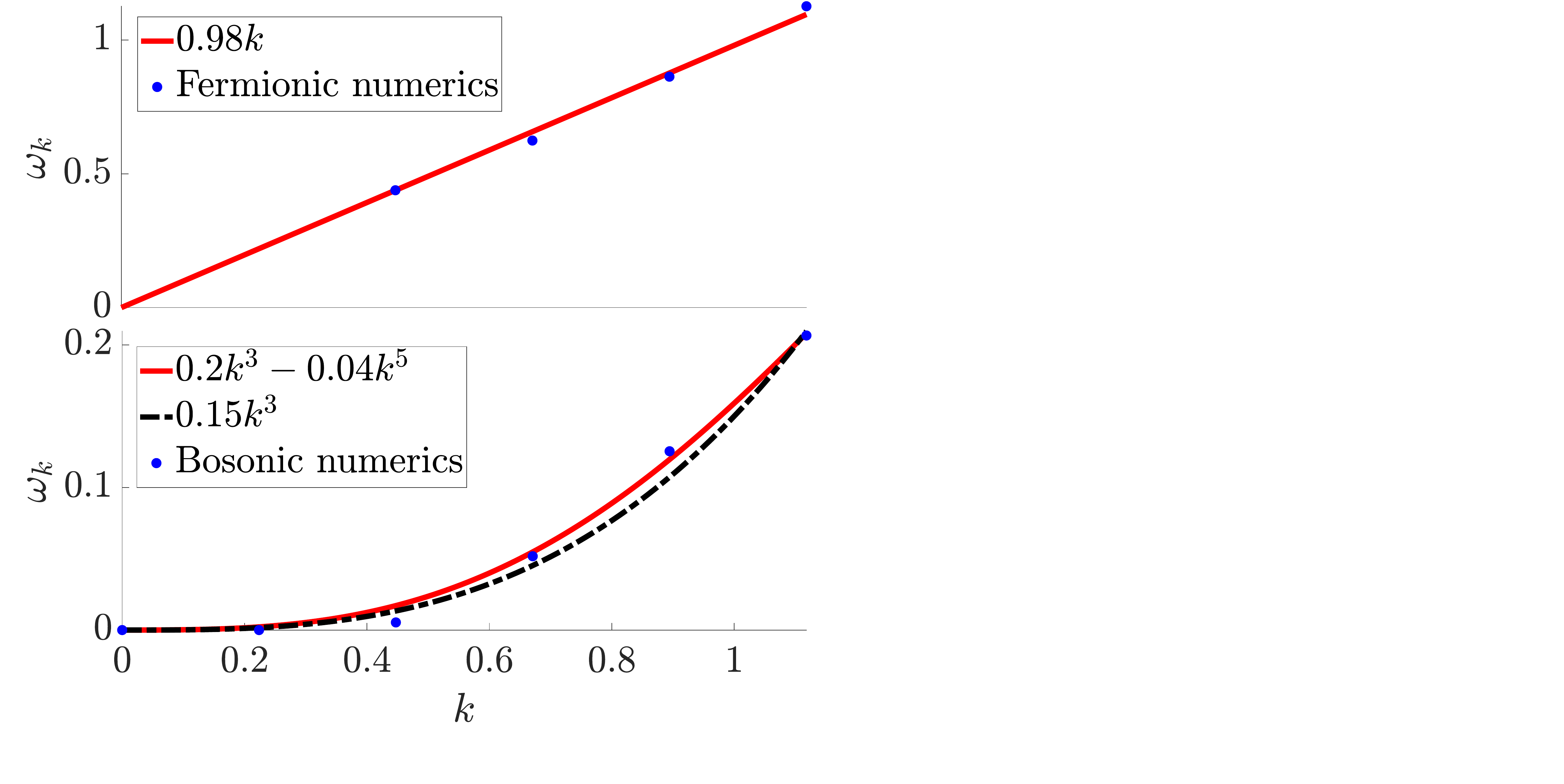}}
	\caption{Same as Fig. \ref{fig:newdisrel}, for the Pfaffian/vacuum interface with $N=10$ bosons. The influence of the curvature corrections here is not as drastic for the bosonic mode as in Fig. \ref{fig:newdisrel} but we still see a better agreement with numerical result. The fermionic mode spectrum (included here for reference) is not affected. \label{fig:newdisrel3}}
\end{figure}

We write the effective edge theory for such system as
\begin{equation}
	S=\int d^2x(i\psi\dot{\psi}-vi\psi\psi')+S_B,
\end{equation}
where $S_B$ is the action given in \cref{eq:originalmodel3}, with different coefficients for $T_0$ etc. which will be estimated by fitting to  numerical data \cref{fig:newdisrel3}. Accordingly, we find that the dispersion relation for pure bosonic excitations to be
\begin{equation}
	\omega\approxeq 0.2 k^3-0.04k^5,
\label{eq:cordisrel1}
\end{equation}
from which we extract $T_0\approxeq0.031$ and $T_2\approxeq-0.0031$. The fermionic excitation dispersion relation remains linear with $v\approxeq 0.98$, all in units of $|V_0|$. Since there is no particle-hole symmetry for bosons, $T_1$ is {\em not} expected to be zero in this case. However it does {\em not} enter the quadratic action (\ref{eq:originalmodel4}), as a result of which it has no effect on the dispersion. It does, however, has effects on the action beyond the quadratic level, which gives rise to boson-boson interaction.

Due to the presence of the fermionic mode, the interface here is more like a "superstring"\cite{Polchinski1998}, although there is no supersymmetry here. It has been shown\cite{PhysRevLett.126.206801}, however, the Moore-Read (Pfaffian) edge can be made supersymmetric, and this supersymmetry can even be broken spontaneously, resulting in a novel Goldstino mode at low energy. It is thus worthwhile to explore the analogy between the dynamics of this interface and superstring in future work.

\section{topological term of the string theory via Chern-Simons theory with fluctuating boundary}
\label{sec: Chern-Simons theory with time-dependent boundary}

So far we have treated interfaces as a string-like objects such that the energy of the system is a functional of its length and local curvature. We showed that such identification enabled us to derive an effective low energy interface theory, that gives a low-energy dispersion relation that agrees with the numerical calculations of the microscopic models extremely well. We have achieved this success by combining the energy functional with a topological term in the action that was derived earlier for edge theory, which, as we discussed in Sec. \ref{subsec: General form of string acion}, is only appropriate for small fluctuations of the interface. In this section we attempt to derive the topological action that is appropriate for large fluctuations, for the simplest case of interface between the fermionic $\nu=1$ QHD and vacuum.

Let us recall how the topological term in the edge theory, namely the first term of Eq. (\ref{eq:originalmodel}) was obtained \cite{Wen2007}. One starts with the bulk Chern-Simons theory, and place it on a 2D manifold with {\em fixed} (or {\em time-independent}) boundary. One finds that the Chern-Simons theory is {\em not} gauge-invariant, as a gauge transformation generates an additional boundary term to the action. This indicates there must be boundary degrees of freedom {\em not} captured by the bulk Chern-Simons theory, which are precisely the edge states. To describe such edge states, one impose certain "gauge fixing" condition at the boundary, which turns certain gauge "degree of freedom" there to physical degree of freedom, whose dynamics is actually sensitive to the gauge fixing condition. One thus has to carefully choose the gauge fixing condition (in an {\em ad hoc} way) to get the correct edge dynamics.

The situation we are facing here has three important differences. (i) The interface, as parameterized by ${\vec{X}}(t, \sigma)$, is fluctuating and {\em time-dependent}. (ii) ${\vec{X}}(t, \sigma)$ {\em is} the explicit physical boundary degree of freedom, so no new degree of freedom needs to be introduced. (iii) Perhaps most importantly, the dynamics of ${\vec{X}}(t, \sigma)$ comes from the energy functional $E[{\vec{X}}(t, \sigma)]$ in (\ref{eq:string action}). We thus should not get (additional) dynamical term from gauge-fixing the Chern-Simons theory. As we show below, these allow us to derive the appropriate topological term in the action (\ref{eq:string action}).

%of the treatment of the boundary as a string at low energy tempts one to describe the boundary  action in terms of its coordinates only, as in a Nambu-GOTO action.
%
%However, this task seems to be rather haunted by subtle problems, related to mapping of the chiral density of the boundary excitations with the length of deviation of the boundary from the equilibrium. Taking this mapping literally even works for low energy excitations, would create inconsistencies when we want to include high energy excitations (in other words deviations from the equilibrium boundary which are not infinitesimal).
%
%If we describe the boundary by the coordinates of the points on the boundary, those coordinates should transform as a scalar under reparametrization transformation since the physical shape of the boundary should be invariant under such transformation. The deviation of the boundary from the equilibrium is given as the derivative of the $\phi$ since $\phi$ transforms as a scalar, than the deviation transforms as a scalar density, thus this creates an inconsistency. One of the peculiar subtleties of the CS theory with boundary is, as the fields of the theory fluctuates, the physical boundary i.e. the geometry of the boundary also fluctuates  because of the mentioned mapping.

%In order to get more insight on dynamic and kinematic nature of the boundary of the quantum hall droplet,
To proceed, we go back to the bulk Chern-Simons theory and now explicitly assume a kinematically time dependent boundary, and discuss its gauge properties. The theory is given as
\begin{equation}
	S_{\text{CS}}=\int_{-\infty}^{\infty} dt\int_{D(t)}d^2x\Big[-\frac{1}{4\pi}\epsilon^{\mu\nu\lambda}a_\mu\partial_\nu a_\lambda+\frac{1}{2\pi}\epsilon^{\mu\nu\lambda}A_\mu\partial_\nu a_\lambda\Big]
\label{eq: Chern-Simons}
\end{equation}
where $A$ is the background gauge field and $D$ is time-dependent region that the droplet lives. Under the gauge transformation $a_\mu\to a_\mu+\partial_\mu\Lambda$ the total change of the action is
\begin{align}
&\Delta S_{\text{CS}}=\int\displaylimits_{R} dt\int\displaylimits_{D(t)}d^2x\epsilon^{ij}\partial_t(\Lambda\partial_ia_j)\nonumber\\{\notag}
&+\int\displaylimits_{R} dt\int\displaylimits_{\partial D(t)}dx^j\Lambda(-\partial_ta_j+\partial_j a_t)\\
&= \int\displaylimits_{R} dt\int\displaylimits_{\partial D(t)}\left[dx^j(-\partial_ta_j+\partial_j a_t)\Lambda - dx^\ell v^k\epsilon_{k\ell} \epsilon^{ij}(\Lambda\partial_ia_j)\right] \label{eq: stringliketheo}
\end{align}
where we have used the
Reynolds-Leibniz rule for differentiation under integral \cite{Flanders1973} and $R$ is $[-\infty,\infty]$, $\vec{v}(\vec{X},t)$ is the velocity vector of the boundary at point $\vec{X}$. Note in earlier work considering {\em fixed} boundary, $\vec{v}=0$ and only the first term above is present, and it is the existence of such boundary term that ruins the gauge invariance of the Chern-Simons theory in manifolds with boundaries. We note, however, in the present case,
we can make $\Delta S_{\text{CS}}$ zero, and hence the theory gauge invariant, if we impose the following {\em local} boundary condition:
\begin{equation}
J_n(\vec{X},t)  = J_0 v_n(\vec{X},t),
\label{eq: boundary_condition}
\end{equation}
where we used the fact that the bulk current\cite{Wen2007}
\begin{equation}
J^\mu=\frac{1}{2\pi}\epsilon^{\mu\nu\lambda}\partial_{\nu}a_{\lambda},
\label{eq: current}
\end{equation}
and the subscript $n$ stands for the normal component of a (spatial) vector at the boundary point $\vec{X}$. In particular $J_0 = \frac{1}{2\pi}$ is the density of the $\nu = 1$ QH liquid which is a constant when the back ground magnetic field $B = \epsilon^{ij} \partial_i A_j$ is uniform, although everything we say apply to non-uniform $B$ as well.

The boundary condition (\cref{eq: boundary_condition}) has a very clear physical meaning: when there is bulk current $J_n$ flowing toward the boundary, the boundary must recede with speed $v_n$ to accommodate the excess charge. As a result charge conservation is preserved, hence gauge invariance. This contrasts with the usual approach to the edge theory which assumes the boundary is fixed, as a result charge would not be conserved (hence lack of gauge invariance, or charge conservation becomes "anomalous" at the boudnary) without the additional degree of freedom $u$ introduced by hand. It should be clear from Fig. \ref{fig:stringgauge} that $u$ is nothing but the fluctuation of the boundary. Our discussion above thus provides a new perspective of the traditional edge theory. A bonus of \cref{eq: boundary_condition} is it also enforces area conservation (which is nothing but charge conservation for an incompressible liquid).

The topological term of the full string theory, $S_{top}$ of (\ref{eq:string action}), is nothing but the Chern-Simons action (\ref{eq: Chern-Simons}), subject to the constraint, \cref{eq: boundary_condition}, which is an (implicit) functional of ${\vec{X}}(t, \sigma)$. In other words $S_{top}$ can be obtained from integrating over the gauges fields $a$ in (\ref{eq: Chern-Simons}), subject to the constraints of \cref{eq: boundary_condition} and \cref{eq: current}.

%%%%%
Armed with \cref{eq: boundary_condition} which couples the boundary (interface) with the bulk Chern-Simons field $a$, we are ready to derive the topological action $S_{top}$ in (\ref{eq:string action}). To do that we adopt the gauge-fixing condition $a_0(\vec{X})=0$ at the interface, which is known {\em not} to generate (addtional) dynamical terms\cite{Wen1992}. Following the standard procedure, we introduce a scalar field $\phi$ such that
\begin{equation}
a_j = A_j + \partial_j \phi
\label{eq: phi}
\end{equation}
in the bulk which solves the bulk constraint $\epsilon^{ij} \partial_i a_j = \epsilon^{ij} \partial_i A_j$ obtained from integrating over $a_0$. With (\ref{eq: phi}) $S_{\text{CS}}$ reduces to a boundary term:
\begin{equation}
	S_{top}=-\frac{1}{4\pi}\int dt \int d\sigma \dot{\phi}\phi'
\label{eq:topological term}
\end{equation}
which is essentially the first term of (\ref{eq:originalmodel}), but with arbitrary parameterization along the boundary ${\vec{X}}(t, \sigma)$, and now $\phi' = \frac{\partial\phi}{\partial\sigma}$ so $S_{top}$ as well as the entire action (\ref{eq:string action}) is explicitly invariant under re-parameterization. Most crucially however, $\phi$ is constrained by (\ref{eq: boundary_condition}), via the identification
\begin{equation}
J_n(\vec{X},t)  = \frac{1}{2\pi} (\partial \dot{\phi}/\partial\sigma)/|\partial{\vec{X}}/\partial\sigma|.
\label{eq: boundary current}
\end{equation}
As a result $S_{top}$ is an (implicit) functional of ${\vec{X}}(t, \sigma)$.

\section{concluding remarks}
\label{sec: concluding remarks}

In this paper we have studied string-like theories for free interfaces separating quantum Hall droplets from vacuum. The two cases we studied correspond to bosonic and superstrings respectively, in terms of the degrees of freedom associated with their interfaces. While they may seem similar to the usual quantum Hall edges, we demonstrated their dynamics is very different when such interfaces are not pinned by confining potentials, and requires string-like theories to describe. We derived such a theory for the case of bosonic string, and leave such a derivation for superstring to future work.

Physically more interesting cases are interfaces between different quantum Hall liquids. A particularly interesting example is the $\nu=5/2$ quantum Hall liquid, where multiple energetically competitive states are present, and may form various interfaces (see \cite{Ken2022} for a recent review). The methods we developed in this paper can be generalized to such interfaces as well. In particular, the energy term in (\ref{eq:string action}) should be present in the generic case. The topological term, on the other hand, needs to be derived by combining the bulk Chern-Simons-like theories on both sides of the interface, supplemented by appropriate boundary conditions like \cref{eq: boundary_condition}.

\section*{Acknowledgements}
We thank Qi Li and Ken Ma for useful correspondences.
This work was supported by the National Science Foundation Grant No. DMR-1932796, and performed at the National High Magnetic Field Laboratory, which is supported by National Science Foundation Cooperative Agreement No. DMR-1644779, and the State of Florida.
\bibliography{ref}

\end{document}